\documentclass[conference]{IEEEtran}
\usepackage{booktabs}
\IEEEoverridecommandlockouts
\usepackage{cite}
\usepackage{amsmath,amssymb,amsfonts}
\usepackage{algorithmic}
\usepackage{svg}
\usepackage{graphicx}
\usepackage{textcomp}
\usepackage{pifont}
\usepackage{float}

\usepackage{pgfplots}
\usepackage{url}
\usepackage{hyperref}
\usepackage{xcolor}
\def\BibTeX{{\rm B\kern-.05em{\sc i\kern-.025em b}\kern-.08em
    T\kern-.1667em\lower.7ex\hbox{E}\kern-.125emX}}
\usepackage{array}

\newboolean{showcomments}
\setboolean{showcomments}{false}
\ifthenelse{\boolean{showcomments}}
{

\newcommand{\mynote}[2]{
    {\color{#1}Note: #2}
}

\newcommand{\note}[2]{
    {\color{#1}Note: #2}
}

\newcommand{\Dirk}[2]{
    {\color{#1}\fbox{\begin{minipage}{0.8\linewidth}Dirk: #2\end{minipage}}}
}

\newcommand{\AuthorX}[2]{
    {\color{#1}\fbox{\begin{minipage}{0.8\linewidth}AuthorX: #2\end{minipage}}}
}

\newcommand{\Yisu}[2]{
    {\color{#1}\fbox{\begin{minipage}{0.8\linewidth}Yisu: #2\end{minipage}}}
}
 
} 
{ 
\newcommand{\mynote}[2]{}
\newcommand{\note}[2]{}
\newcommand{\Dirk}[2]{}
\newcommand{\Yisu}[2]{}
\newcommand{\AuthorX}[2]{}
} 

\newcommand{\dirk}[1]{\Dirk{blue}{#1}}

\begin{document}

\title{Affordable HPC:\\
Leveraging Small Clusters\\
for Big Data and Graph Computing}


\author{\IEEEauthorblockN{1\textsuperscript{st} Ruilong WU}
\IEEEauthorblockA{\textit{IOT Thrust, INFO Hub} \\
\textit{HKUST(GZ)}\\
Guangzhou, China \\
rwu408@connect.hkust-gz.edu.cn}
\and
\IEEEauthorblockN{2\textsuperscript{nd} Yisu Wang}
\IEEEauthorblockA{\textit{IOT Thrust, INFO Hub} \\
\textit{HKUST(GZ)}\\
Guangzhou, China \\
ywang418@connect.hkust-gz.edu.cn}
\and
\IEEEauthorblockN{3\textsuperscript{rd} Dirk KUTSCHER*}
\IEEEauthorblockA{\textit{IOT Thrust, INFO Hub} \\
\textit{HKUST(GZ)}\\
Guangzhou, China \\
dku@hkust-gz.edu.cn}

}

\maketitle

\begin{abstract}
This study explores strategies for academic researchers to optimize computational resources within limited budgets, focusing on building small, efficient computing clusters. It delves into the comparative costs of purchasing versus renting servers, guided by market research and economic theories on tiered pricing. The paper offers detailed insights into the selection and assembly of hardware components such as CPUs, GPUs, and motherboards tailored to specific research needs. It introduces innovative methods to mitigate the performance issues caused by PCIe switch bandwidth limitations in order to enhance GPU task scheduling. Furthermore,  a Graph Neural Network (GNN) framework is proposed to analyze and optimize parallelism in computing networks.
\end{abstract}


\section{Introduction}
\dirk{Some idea for motivating this: Large ML models, such as LLMs, are becoming more and more powerful and available to end-users (such as Llama-3.1). As they become more powerful, their memory and inference computation requirements exceed the capabilities of inidividual PCs and servers. In order to enable users, research groups etc., to use and experiment with these models, distributed architectures are needed...}

Large machine learning (ML) models, such as language models (LLMs), are becoming increasingly powerful and gradually accessible to end users. However, the growth in the capabilities of these models has led to memory and inference computation demands exceeding those of personal computers and servers. To enable users, research teams, and others to utilize and experiment with these models, a distributed architecture is essential.

In recent years, scientific research has shifted from a "wisdom paradigm" to a "resource paradigm." As the number of researchers and the depth of scientific exploration increase, a significant portion of research computing tasks has moved to servers. This shift has been facilitated by the development of computing frameworks and widespread use of computers, leading to an increased demand for computer procurement.

Despite the abundance of online tutorials for assembling personal computers, information on the establishment of large clusters is relatively scarce. Large Internet companies and multinational corporations usually employ professional architects and engineers or work closely with vendors to optimize their cluster performance. However, researchers often do not have access to these technical details and must rely on packaged solutions from service providers to build  small clusters.

In this study, we aim to bridge this gap by providing opportunities for researchers with limited funds to build small clusters from scratch. We compiled the necessary technical details and guidelines to enable researchers to assemble clusters independently. In addition, we propose a method to mitigate the performance degradation caused by the bandwidth limitations of PCIe switches, which can help researchers prioritize GPU training tasks effectively.

The main contributions of this paper are as follows:

\paragraph{Technical Detail Compilation}We provide a comprehensive guide for researchers with limited funds, helping them to independently build small clusters and contribute to the development of large models.

\paragraph{Performance Optimization}We propose a method to address the performance degradation caused by PCIe switch bandwidth limitations. This method allows researchers to prioritize GPU training tasks effectively, thereby improving the overall cluster performance.

\paragraph{GNN for Network and Neural network parallelism}We propose a GNN (Graph Neural Network) framework that combines neural networks with parallel network flows in distributed systems. Our aim is to integrate different types of data flows, communication patterns, and computational tasks, thereby providing a novel perspective for evaluating the performance of distributed systems.

The remainder of this paper is organized as follows: Section 3 discusses CPU selection, Section 4 discusses GPU selection, Section 5 discusses motherboard selection, Section 6 discusses internal topology, Section 7 discusses network equipment, and Section 8 discusses storage. Section 9 proposes the concept of combining neural networks with distributed network flows. Section 10 provides an outlook for future research.\dirk{need to add section 9 and 10}

\dirk{The main contributions of this paper are...}

\dirk{The remainder of this paper is structured as follows...}

\section{Motivation}

\subsection{Survey on different cloud server rental prices}

Table 1 lists the average research funding per capita for different levels of universities in China as well as general program funded by the National Natural Science Foundation of China. Table 2 surveys the monthly rental prices of a common configuration (96 vCPUs, 192 GB memory, and 1024 GB storage) from the largest cloud service providers, along with the duration these funds can cover rental costs. Table 3 lists the monthly rental prices for an 8-card A100 GPU server from cloud service providers and shows how long these funds can cover rental costs. Additionally, our research on the domestic computing power rental market in China reveals that GPU resources are in such short supply that most vendors are currently sold out.

\begin{table}[ht]
\centering
\caption{Per Capita Funding Categories} 
\begin{tabular}{>{\centering\arraybackslash}p{4.5cm} >{\centering\arraybackslash}p{2.5cm}}
\toprule
\textbf{Category} & \textbf{Per Capita Funding (10,000 USD)} \\
\midrule
General program of NSFC\cite{b2} & 6.82 \\
“211” and Provincial-Ministerial Joint Construction Universities\cite{b1} & 6.14 \\
Ordinary Undergraduate\cite{b1} & 1.84 \\
Junior College\cite{b1} & 0.38 \\
\bottomrule
\end{tabular}
\label{tab:funding_categories}
\end{table}

\begin{table*}[ht]
\centering
\caption{Comparison of NSFC Funding and Cloud CPU Server (96 vCPUs, 192 GB memory, and 1024 GB storage) Prices}
\label{tab:funding_vs_cloud}
\begin{tabular}{p{2.2cm}p{1.5cm}<{\centering}p{2.5cm}<{\centering}p{2.7cm}<{\centering}p{2.7cm}<{\centering}p{2.5cm}<{\centering}}
\toprule
\textbf{Provider} & \textbf{Month Price (USD)} & \textbf{NSFC Gen. Fund/Price} & \textbf{"211" Univ. Fund/Price} & \textbf{Gen. Undergrad Fund/Price} & \textbf{Voc. College Fund/Price} \\
\midrule
GCP & 3751.82 & 18.18 & 16.37 & 4.90 & 1.01 \\
AWS & 3714.24 & 18.36 & 16.53 & 4.95 & 1.02 \\
Azure & 3055.20 & 22.32 & 20.10 & 6.02 & 1.24 \\
Tencent & 1710.34 & 39.87 & 35.89 & 10.76 & 2.22 \\
Alibaba & 1507.32 & 45.25 & 40.74 & 12.21 & 2.52 \\
Huawei & 2509.57 & 27.18 & 24.46 & 7.33 & 1.51 \\
\bottomrule
\end{tabular}
\end{table*}

\begin{table*}[ht]
\centering
\caption{Comparison of NSFC Funding and GPU Cloud Server (8 A100 GPU) Prices}
\label{tab:funding_vs_gpu_cloud}
\begin{tabular}{p{2.2cm}p{1.5cm}<{\centering}p{2.5cm}<{\centering}p{2.7cm}<{\centering}p{2.7cm}<{\centering}p{2.5cm}<{\centering}}
\toprule
\textbf{Provider} & \textbf{Price (USD)} & \textbf{NSFC Funding/Price} & \textbf{"211" Univ. Funding/Price} & \textbf{Undergrad Inst. Funding/Price} & \textbf{Voc. College Funding/Price} \\
\midrule
GCP & 9529.29 & 7.15 & 6.44 & 1.93 & 0.40 \\
AWS & 15054.79 & 4.53 & 4.08 & 1.22 & 0.25 \\
Azure & 16512.41 & 4.13 & 3.72 & 1.11 & 0.23 \\
\bottomrule
\end{tabular}
\end{table*}

\subsection{Theory of Tiered Pricing}
Based on the theory of tiered pricing\cite{b3} \dirk{reference?}, each intermediary in the supply chain adds markups to  product or service prices, according to the cost-plus principle. This markup covers operational costs, generates profits, compensates for risks, and adds value, resulting in a progressive increase in the final price. For researchers with limited funding, directly purchasing computing servers rather than renting them can avoid cumulative price increases from intermediaries, thereby utilizing the budget more effectively and obtaining more computing power.

By directly purchasing computing servers, researchers can bypass the markup applied by rental service providers, who typically incorporate a cost-plus margin into their service offerings. Although purchasing computing servers may require a significant initial investment, it can offer a more cost-effective solution in the long term. Owning their own servers allows researchers to avoid rental fees and to freely configure and manage resources according to their needs. This provides better control over the use of computing resources and eliminates concerns regarding potential price adjustments or usage restrictions imposed by rental service providers.

\section{CPU Selection}
When selecting CPUs, it is essential not only to focus on parameters, such as Instructions Per Cycle (IPC), but also to consider the CPU's topology. As CPU performance improves, IPC enhancement has gradually reached a bottleneck. Increasing IPC often involves expanding the issue width. The issue width of a CPU refers to the number of instructions that the processor can issue to the execution units from the instruction queue simultaneously per clock cycle. This is a crucial measure of the processor's parallel processing capability, directly affecting its throughput and performance.

For example, the latest Apple Silicon chips have a large issue width, with  large cores of up to 8. However, further increasing the issue width yields limited performance gains. Consequently, modern CPUs extensively use chiplet technology, an integrated circuit design method that breaks down complex systems into smaller chip modules. These modules are interconnected using high-speed interconnect technology to form a complete system, achieving higher performance while maintaining production yields.

This modular approach inherently introduces greater latency, altering computational dynamics at the task level. Therefore, when building small clusters, it is critical to consider these factors to ensure that the chosen hardware aligns with the specific computational requirements of the research tasks.

\subsection{CPU internal structure}

\dirk{some intro sentence}
Different CPUs provide distinct internal structures, and these internal structures, along with other CPU parameters, form the basis for our selection.
\subsubsection{AMD EPYC}
\begin{figure*}
    \centering
    \includegraphics[width=0.75\textwidth]{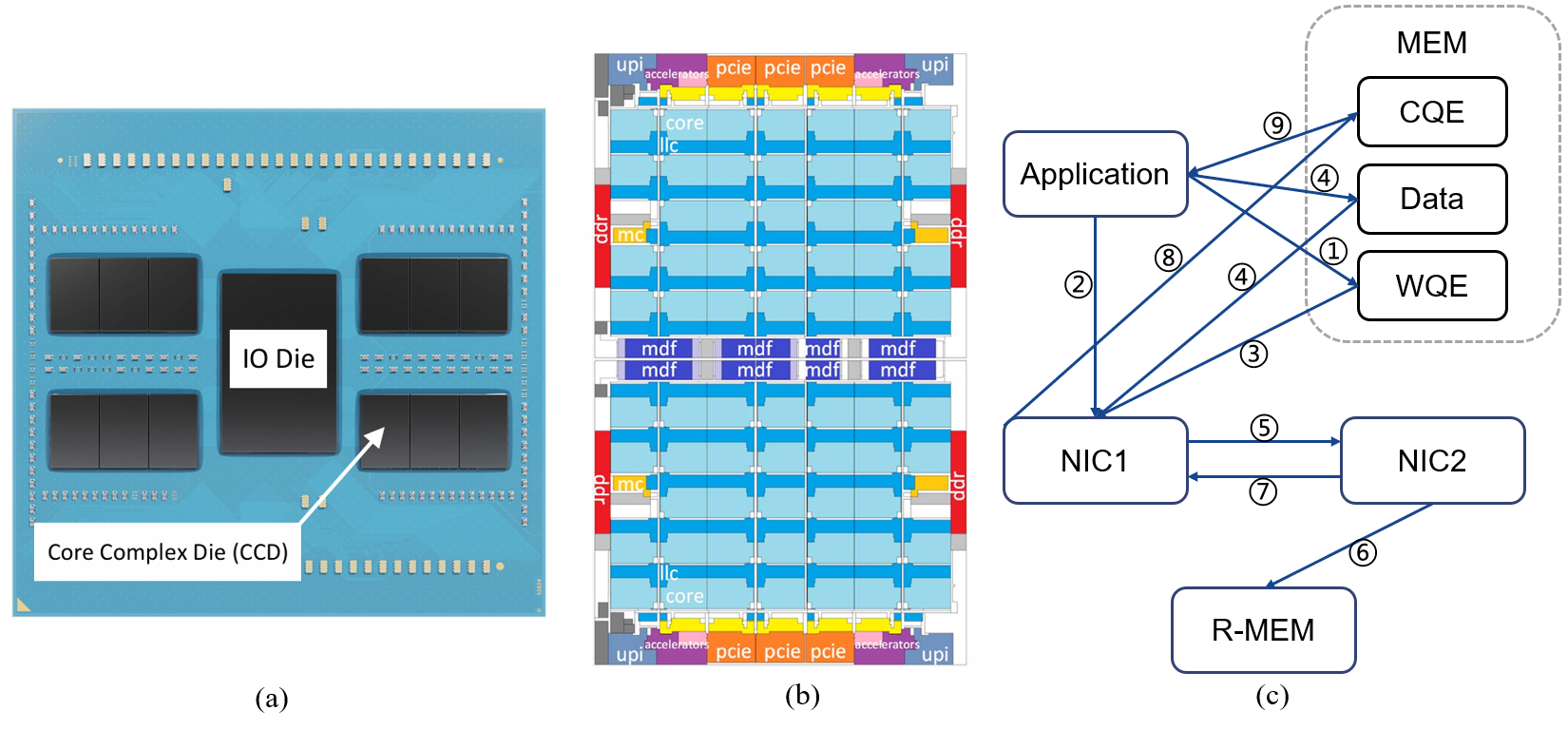}
    \caption{(a)AMD EPYC 9004 configuration with 12 Core Complex Dies (CCD) surrounding a central I/O Die (IOD)\cite{b4} (b)Processor floorplan diagram for 2-die XCC configuration\cite{b5} (c)Standard RDMA over PCIe Transfer Process: \ding{172} Generate Work Queue Element \ding{173} Issue Doorbell \ding{174}Network Card Fetches Task \ding{175} DMA Data to Network Card \ding{176} Data Encapsulation and Transmission \ding{177} Processing at Receiving End \ding{178} Return Completion Message \ding{179} Generate Completion Queue Element \ding{179}Application Polls CQE}
    \label{f1}
\end{figure*}

As shown in the figure\ref{f1}a, the AMD EPYC 9004 series processors have two key components: the core complex die (CCD) and the I/O die (IOD). The CCD contains cores based on the Zen 4 architecture, with each CCD housing up to two core complex (CCX) units. Each CCX can have up to eight cores that share an L3 cache. The IOD acts as the central hub, connecting all CCDs using Infinity Fabric interconnect technology. It also supports up to 12 DDR5 memory channels and up to 128 PCIe Gen5 lanes.

\subsubsection{Intel Xeon}
The Intel Emerald Rapids processor features a unique architecture with two dies, Xeon Node Controller (XNC) and Xeon Scalable Core (XSC), connected via Embedded Multi-Die Interconnect Bridges (EMIBs). This design facilitates high-bandwidth, low-latency communications between the dies, thereby enhancing overall performance and efficiency. Unlike the AMD EPYC series processors, which incorporate an I/O Die (IOD) to handle I/O operations separately, the Intel Emerald Rapids processor integrates PCIe and memory connections directly to the CPU cores. This direct connection can potentially reduce latency and improve data throughput by eliminating intermediary IOD, streamlining data paths within the processor.

\subsubsection{Influence of Topology}
Remote Direct Memory Access (RDMA) is an efficient data transfer technology that allows devices to read from or write to another device's memory directly, without operating system intervention. This approach significantly reduces data transfer latency and CPU overhead, enhancing system performance. It is particularly useful in high-performance computing and large-scale data center applications.

The differences in RDMA performance between the AMD EPYC and Intel Xeon processors primarily stem from their architectural designs. AMD EPYC processors use an I/O Die (IOD) design that centralizes the memory and PCIe controllers into a single module. This design results in a shorter data transfer path from the memory to PCIe, leading to lower latency and higher bandwidth utilization. Conversely, Intel Xeon processors employ a traditional monolithic design in which the memory and PCIe controllers are distributed across different CPU cores. This results in longer data transfer paths that pass through CPU cores, leading to higher latency and lower performance in high-frequency data transfer scenarios.

\subsubsection{Chiplet Placement}
The latest chiplet technology has demonstrated advantages in terms of cost control. Manufacturers can tolerate higher defect rates by masking the damaged parts to produce lower-end products. For example, lower-end AMD EPYC series CPUs may provide only 1-2 cores per chiplet. This design allows each core to have access to the full L3 cache but also results in increased communication time between cores. When selecting a specific model, it is crucial to consider the task requirements: for CPU parallel computing tasks, inter-core communication performance is vital, whereas for GPU-centric tasks, the topology between GPUs is more important.

Once a model is selected, optimization strategies can be developed based on the core topology. For CPU-centric tasks where each chiplet has only 1-2 cores, reducing the frequency of inter-core communication and allocating more time to computation can enhance the performance. Conversely, for GPU-centric tasks, specifying which cores to use and minimizing inter-core communication can improve efficiency. These strategies can help maximize performance and enhance task-processing efficiency, given existing hardware conditions.

\subsection{Other parameters}
\subsubsection{Power}
Manufacturers often split CPUs with the same core count into different models, and determining their specific performance is typically done by examining their Thermal Design Power (TDP). TDP represents the maximum heat dissipation capacity required for the CPU when running high-intensity tasks, and is generally correlated with CPU performance. When selecting an appropriate CPU model, TDP can be used to gauge its performance, allowing for the selection of a model that fits specific task requirements.

For example, for tasks that require high-performance computing, choosing a model with a higher TDP is beneficial, because these models typically have higher frequencies and larger caches, offering greater computational power. However, for applications with higher energy efficiency demands, a model with a lower TDP would be more suitable, as it reduces power consumption and cooling requirements. By comparing the TDP of different models, it is possible to select a CPU that meets the specific task requirements, optimizing the balance between performance and power consumption.
\subsubsection{3D-v cache}
3D Vertical Cache (3D-V Cache) is a technology that increases processor cache capacity and bandwidth through vertical stacking, significantly enhancing system performance and energy efficiency. By utilizing vertical stacking, it achieves greater cache capacity within a limited chip area, reducing data transfer latency and optimizing space utilization. This technology is particularly well-suited for high-performance computing and data-intensive applications.

\subsection{Tools for CPUs}
CPU manufacturers typically provide tools to help users optimize CPU performance, including configuring topology logic, optimizing latency, and setting frequencies. For instance, AMD EPYC processors offer tools that can reduce system latency by adjusting the hardware configurations and BIOS settings. In addition, the Linux TuneD tool\cite{b6} can be used for system-level optimization. These tools and scripts, such as `sysjitter`, can measure and report jitter events, aiding in the optimization of server performance to meet the stringent low-latency requirements of applications like high-frequency trading.

\section{GPU Selection}
\begin{figure*}
    \centering
    \includegraphics[width=1\textwidth]{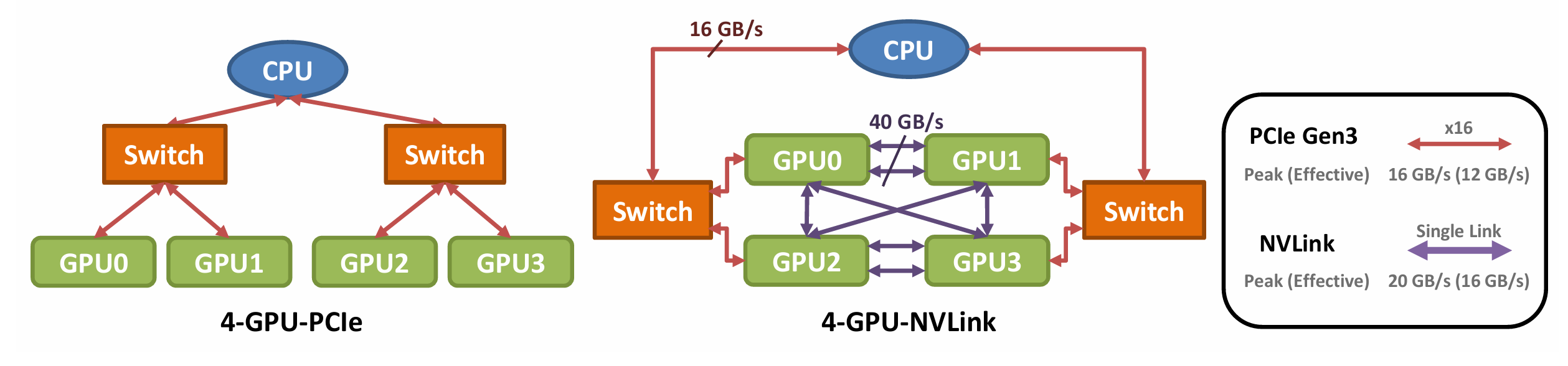}
    \caption{Comparing 4-GPU topologies with NVLink and PCIe. In 4-GPU-NVLink, GPU0 and GPU1 have 40 GB/s peak bandwidth between them, as do GPU2 and GPU3. The other peer-to-peer connections have 20 GB/s peak bandwidth\cite{b12}}
    \label{fig:enter-label}
\end{figure*}
\subsection{Memory And Cache}
The primary purpose of building small clusters for currently popular large model applications is to address the issue of insufficient GPU memory. Several frameworks have been developed to optimize GPU memory usage, such as flash attention. These frameworks have now been expanded to leverage the GPU's L2 cache, utilizing  different levels of cache speed and capacity to achieve fast data transfer and retrieval. This approach accelerates the training and inference of large language models.

\subsection{Topology}
Typically, NVIDIA GPUs are interconnected using NVLink, which can be configured in two different topologies: trees and point-to-point. Tree Topology uses a central node to coordinate communication between GPUs, making it easy to manage and well-suited for applications requiring frequent CPU communication and high scalability, such as data analytics. On the other hand, the Point-to-Point Topology offers high bandwidth and low latency direct GPU connections, ideal for high-performance computing and deep learning tasks that require extensive parallel computation and low-latency data transfer. Choosing the appropriate topology can fully leverage the advantages of NVLink, enhancing system performance and efficiency.

\subsection{Compatibility between different GPUs}
When performing computations on GPUs, a task must be assigned to GPUs of the same model. However, recent frameworks have enabled computations  across different architectures from the same manufacturer. For example, Metis\cite{b7} can run the same task on both NVIDIA's T4 and V100 GPUs, whereas the SCALE\cite{b8} framework supports running CUDA programs directly on AMD GPUs. These emerging frameworks are significant for reducing cluster costs, as they allow for a more flexible utilization of existing hardware resources and improve computational efficiency.

\section{Motherboards}
When selecting a motherboard, the first consideration should be its power-delivery capability. Some motherboards may not be able to support the maximum CPU performance because they cannot provide sufficient power. Next, the number of PCIe slots should be examined, as some motherboards may not expose all PCIe lanes available from the CPU. Additionally, different motherboards offer varying numbers of memory slots, which is crucial for applications that require large memory capacities.

For most server CPU motherboards,  Intelligent Platform Management Interface (IPMI) functionality is essential, because it greatly facilitates user management. However, to reduce costs, some motherboards might not include the IPMI functionality. In such cases, external hardware, such as PiKVM\cite{b9}, TinyPilot\cite{b10}, and ONEKVM\cite{b11}, can be used to emulate the IPMI functionality.

\section{Intra-topology}
\subsection{Pcie switch}
In practice, we observed that our current setup relies heavily on PCIe switches in the internal topology. Existing PCIe switches typically have 16 upstream lanes and 24 to 144 downstream lanes. This topology is highly effective in an 8-card SXM architecture, as the interconnect bandwidth between GPUs is sufficient, whereas the resources for the CPU-GPU interconnect are minimal. However, this approach is inadequate for multi-tenant cloud service scenarios because the low upstream bandwidth severely hampers task allocation speed of the scheduler, leading to low GPU utilization. Furthermore, this fixed topology can lead researchers to overlook the impact of PCIe switches on performance. Our data transfer time measurements indicated that PCIe switches are unnecessary when the demand for PCIe lanes is low.
\begin{figure*}
    \centering
    \includegraphics[width=1\textwidth]{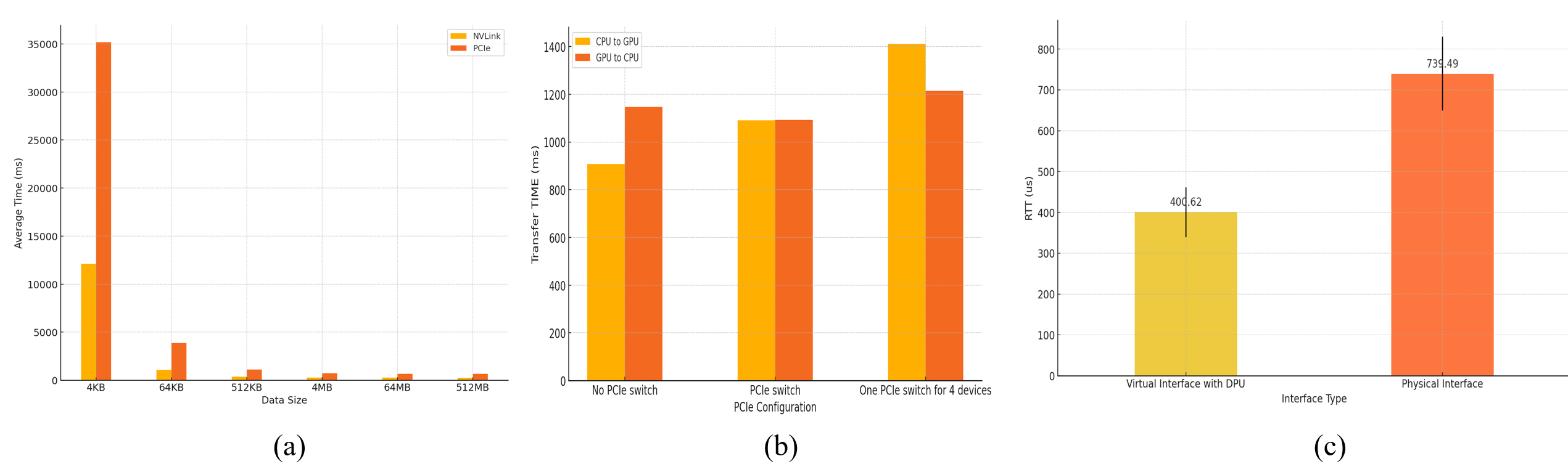}
    \caption{(a)Average Time to Transfer 10GB of Data between GPUs (b)Average time to transfer 10GB of data (c)Mean RTT Between DPU and Tradition Method}
    \label{f3}
\end{figure*}

To address the issue of multiple GPUs transmitting under the same PCIe switch, we designed a method to insert intervals between parallel tasks. This approach staggers asynchronous operations during PCIe transfers, thereby reducing the strain on the CPU. By optimizing transfer scheduling, this method decreases the CPU load and increases GPU utilization. This optimization is particularly crucial in multi-tenant environments, where it can significantly enhance overall system performance and resource utilization.
\begin{figure}
    \centering
    \includegraphics[width=1\linewidth]{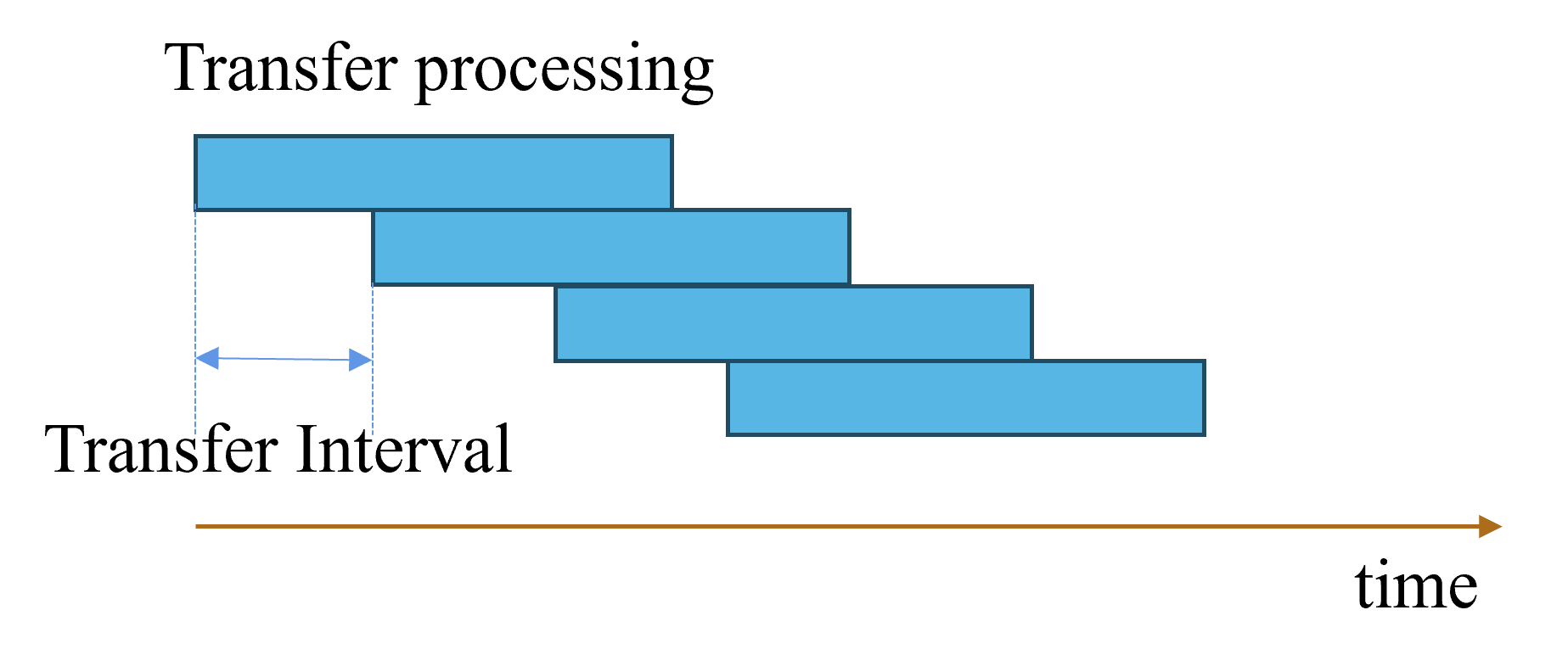}
    \caption{Our design}
    \label{fig:enter-label}
\end{figure}

Despite the uneven cache latency introduced by PCIe, which adds extra transmission time, it also results in some unexpected benefits. Because of this characteristic, we can assume that the transmission time for a PCIe x16 lane is less than double that of a PCIe x8 lane. This is because fixed transmission operations inevitably demand more time than actually required. We assume that the transmission time for the same data consists of two parts: \(a + b\), where \(a\) is the fixed transmission operation time, and \(b\) is the extra operation time introduced by PCIe cache inconsistency. When the transmission bandwidth is reduced owing to a PCIe switch or PCIe bifurcation, the actual operation time becomes \(2a + b\), which is less than twice the transmission time.
Based on this, we can utilize PCIe switches or PCIe bifurcation technology to increase the number of GPUs, thereby enhancing CPU utilization. In multi-GPU setups, these technologies allow the connection of more GPUs without significantly increasing the transmission time, thus improving  parallel computing capabilities and resource utilization of the system. This method not only optimizes the transmission efficiency but also effectively enhances the overall system performance. In high-performance computing and deep learning applications, which require extensive parallel computation, this approach can significantly boost computational efficiency and resource utilization.

\subsection{Socket direct}
In traditional distributed architectures, network card connections are typically arranged, as shown in the accompanying diagram, where the network card is connected directly to CPU1. Although the PCIe connection offers very high speeds, the connection from CPU2 to CPU1's network card still requires passing through the UPI interface. Our tests on cross-CPU data transmission times confirmed that crossing CPUs increases the execution time. NVIDIA introduced the Socket Direct technology, which allows both CPU1 and CPU2's PCIe interfaces to be directly linked to the network card. In this setup, the location of the network card is equidistant from both CPU1 and CPU2, yet many current studies have failed to consider this arrangement. This oversight can lead to inefficiencies in network communication, particularly in systems in which equal access to network resources is critical for performance optimization.

\begin{figure}
    \centering
    \includegraphics[width=1\linewidth]{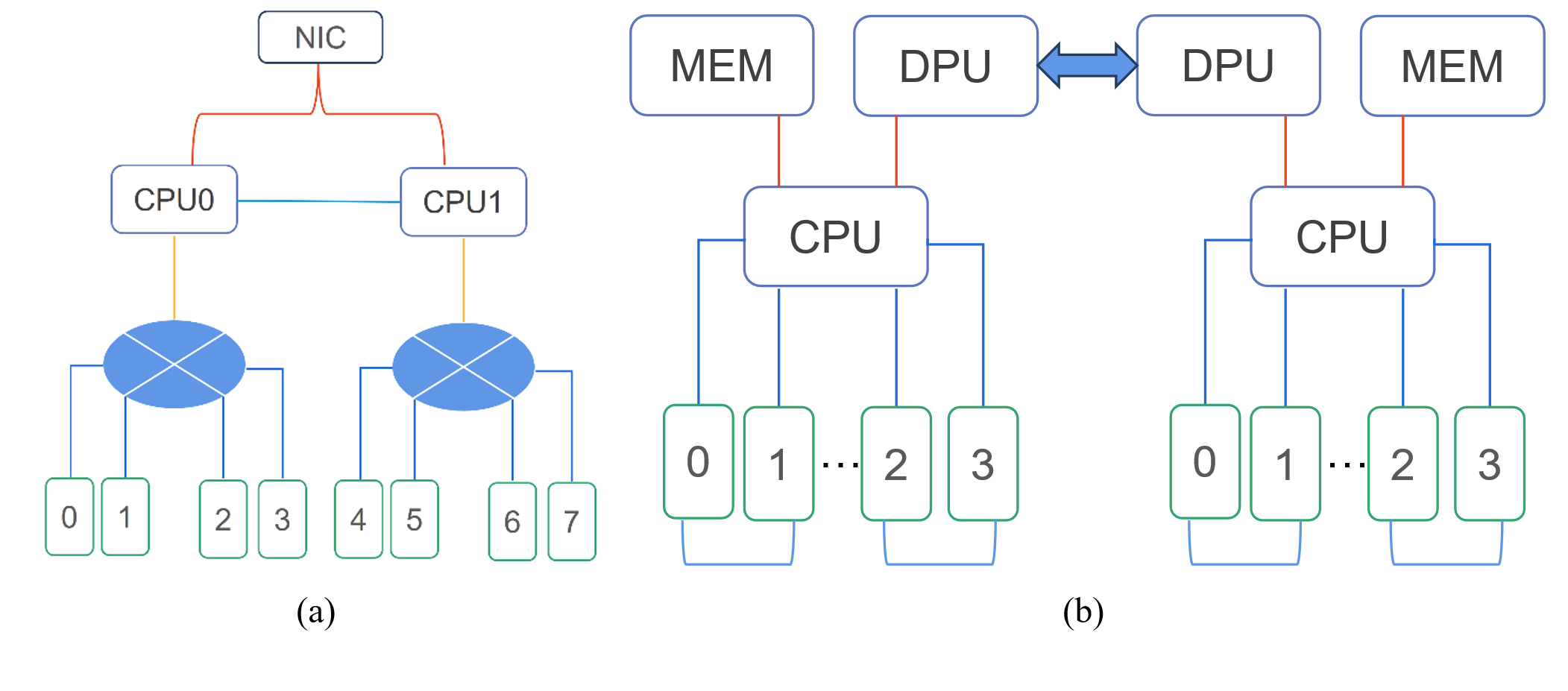}
    \caption{(a)Socket Direct (b)GDR(GPU Direct Remote Direct Memory Access)}
    \label{fig:enter-label}
\end{figure}

\subsection{GDR(GPU Direct Remote Direct Memory Access)}
Traditional  Remote Direct Memory Access (RDMA) technology allows for the remote reading of a CPU's memory; however, when it comes to manipulating GPU memory from a remote location, the data typically needs to be transferred to the CPU's memory first. This step can create bottlenecks when exchanging data among GPUs across different machines. However, if it were possible to establish remote memory addresses directly on GPUs, this could facilitate direct memory transfers from one GPU to another, substantially alleviating data exchange issues between GPUs on different machines.

NVIDIA's BlueField  Data Processing Unit (DPU) cards are equipped to enable this functionality. These DPUs can manage direct remote memory access operations, including those involving GPU memory, thereby streamlining the process of moving data directly between GPUs, without the intermediary step of routing through the memory of the host CPU. This capability is particularly crucial in environments where GPUs are leveraged for high-performance computing and machine learning tasks that require efficient inter-GPU communication.

\subsection{Device manufacturer settings}
In practice, we observed a variability in the internal topology of the machines. Server manufacturers build multiple data paths within servers to meet different needs, providing users with options for which path to use. This results in different internal topologies even within the same server model, making topology optimization closely tied to the specific tasks being executed. This variability can affect the training time of multi-task deep neural network (DNN) training tasks.

To address this issue, we propose a method that leverages natural language processing (NLP) to optimize internal topology. By analyzing server manuals and combining this information with the current topology, we can select specific internal topologies for the servers. This approach allows for the automatic selection of the optimal internal data path in diverse server architectures, thereby optimizing the task execution efficiency.

Additionally, the heterogeneity of the current servers further complicates topology selection. For example, the popular HPC scheduling command Slurm allocates tasks across different nodes, increasing the granularity of resource distribution. In such cases, multi-GPU computation is restricted to the selected task nodes allocated by the scheduler, significantly limiting the task flexibility.
\begin{figure}
    \centering
    \includegraphics[width=1\linewidth]{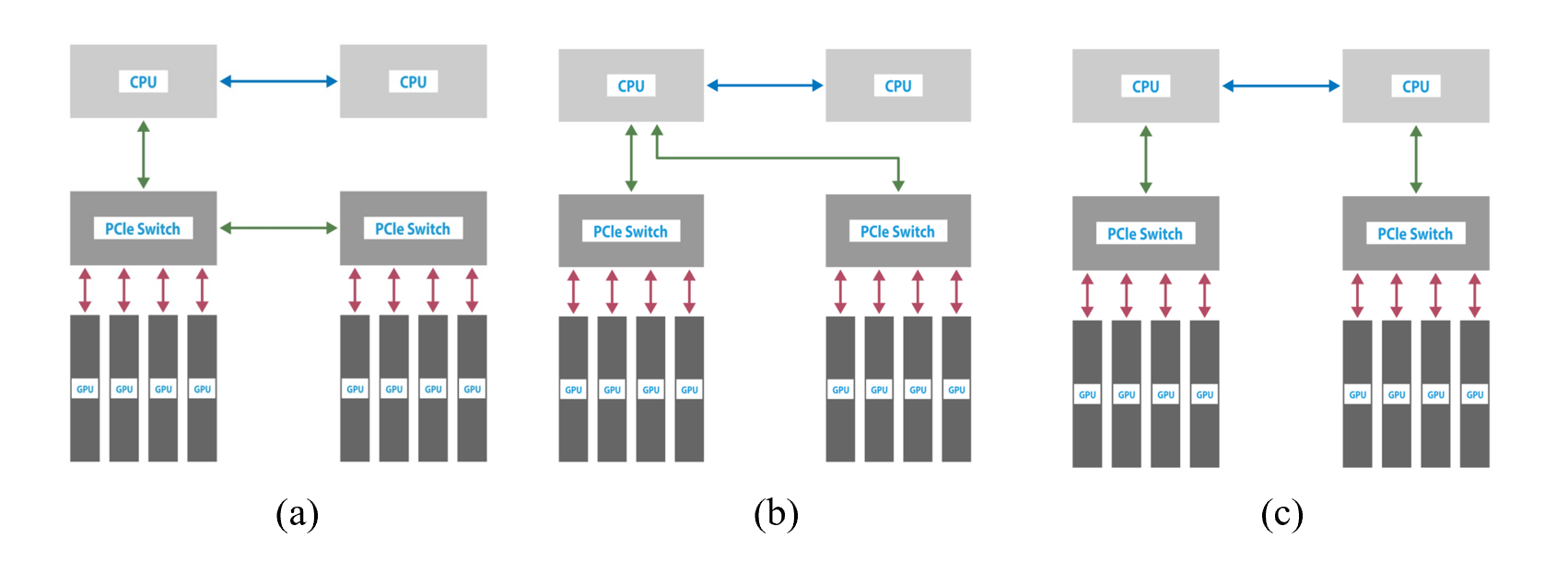}
    \caption{Different models of the same server (e.g., R5300 G5)\cite{b13}}
    \label{fig:enter-label}
\end{figure}

\section{Network}

\dirk{some intro sentence}
In an interconnected network of distributed clusters, some new types of hardware are available for selection.

\subsection{Programmable Switch}
Programmable switches enable flexible definition of packet processing behaviors through high-level programming languages. P4 is currently the dominant programming language used for this purpose, allowing developers to specify how packets are processed and forwarded. The Portable Switch Architecture (PSA) is the current standard architecture for programming packet-level processing and forwarding in P4 switches.\dirk{this is a bit too much high-level marketing. Maybe we call the subsection "Programmable Switches" and then say that P4 is the current dominant programming language, and the Portable Switch Architecture (PSA) is the current architecture for programming packet-level processing and forwarding. The whole approach {\em is} SDN, so we don't have to say that it's suitable for SDN and NFV.}

In distributed deep neural network (DNN) training, P4 switches demonstrate significant advantages. \dirk{"can provide some advantages"?} For example, SwitchML\cite{b14} leverages a programmable data plane to accelerate distributed parallel training by using high-bandwidth programmable switches to scale the maximum bandwidth of the parameter server to the terabyte level, ensuring scalability. ATP\cite{b15} uses programmable switches to dynamically perform gradient aggregation across multiple racks in a data center, thereby optimizing the overall efficiency of the training process. ESA\cite{b16} optimizes switch memory usage through pre-allocation and priority scheduling, significantly increasing switch memory utilization and further reducing job completion time (JCT).

Although solutions such as SwitchML and ATP have effectively addressed many challenges faced by parameter servers in distributed machine learning training, several inherent issues with programmable switches  limit their widespread adoption for foundational DNN training. The primary concerns are as follows:

\subsubsection{Limited Memory Capacity}
Programmable switches typically provide limited memory, which constrains their ability to simultaneously handle large amounts of data. For example, in SwitchML, each worker can have only four outstanding packets at any time to match the slots in the switch, with each packet being 1100 bytes. This limitation affects the efficiency and scalability of operations, such as aggregation.

\subsubsection{Lack of Advanced Functionalities}
The absence of complex functionalities in programmable switches severely restricts their ability to implement sophisticated features such as optimizers. In the case of SwitchML, workers must perform quantization before aggregation, further increasing the Job Completion Time (JCT).

\subsubsection{Cost Issues}
The high cost of programmable switches, which can reach up to \$50,000 each, poses a significant barrier to their widespread adoption. For small to medium-sized data centers and companies, the expenditure required to purchase programmable switches for training acceleration can be prohibitive. This budget could alternatively be allocated to more versatile computing resources, such as CPUs and GPUs, which can handle a broader range of tasks.

\subsubsection{Future Support Concerns}
The programmable switch community is still in its developmental phase and requires broader task migration to establish its position. However, due to market uncertainties, the development and deployment of programmable switches have recently slowed. Following Intel's acquisition of Barefoot Networks, the strategic focus has shifted away from advancing programmable switch projects\cite{b17}, raising doubts about the future support and development of this technology.

\subsection{DPU(Data Processing unit)}
A Data Processing Unit (DPU) is a programmable processor designed specifically for handling network and storage data streams. It integrates high-performance computing cores, network interfaces, storage interfaces, and acceleration engines to offload and accelerate network, storage, and security tasks from the server, thereby enhancing overall system performance and efficiency. DPUs are commonly used in data centers and high-performance computing environments to optimize and accelerate data processing tasks while reducing the load on the main CPU, improving the overall performance and reliability of the system.

\subsubsection{Precision Clock}
DPUs feature precise clock capabilities that can reduce time differences between nodes to the nanosecond level. Compared with traditional millisecond-level synchronization algorithms, this improvement significantly increases the transmission speed and reduces jitter during data transfer. This precision is particularly crucial in distributed deep neural network (DNN) training, ensuring that nodes perform synchronous operations at exact times, thereby enhancing training efficiency and model consistency.

\subsubsection{SNAP}
The SNAP functionality of DPUs allows the virtualization of local drives, connecting them to backend storage systems over the network. This reduces the number of hard drives installed on the PCIe lanes of the system, freeing up more PCIe lanes to accommodate additional GPUs. By virtualizing storage resources, DPUs not only increase the storage access speed but also optimize resource utilization, improving overall system scalability and flexibility.

\subsubsection{Latency}
As shown in figure \ref{f3}c, the DPU connects to the host via PCIe lanes, significantly reducing the communication time compared with traditional multi-node parameter servers. Furthermore, DPUs support GDR (GPUDirect Remote Direct Memory Access), allowing them to  read data directly from the GPU, eliminating the overhead of reading from the host memory. This reduces the latency and enhances the performance. This architectural optimization ensures rapid data transfer and processing, significantly improving the overall efficiency of the distributed training.

\section{Storage}
Vendors often recommend purchasing a switch when configuring storage for a small cluster. However, the high cost of switches can be prohibitive to many users. In reality, if there is no frequent need for interconnectivity, such as running the same task on all machines simultaneously, a switch is entirely unnecessary. We can leverage new technologies available on network cards to achieve functionality similar to that of a switch.

For example, Intel's E810 network card supports NIC partitioning, allowing users to split a 100Gb port into four 25Gb ports. By connecting the 100Gb port to the storage device, there is no need to purchase a switch to meet the basic storage requirements. This approach not only saves costs but also effectively utilizes existing network hardware resources.

With this configuration, users can flexibly allocate bandwidth without sacrificing performance, thereby achieving an efficient data transfer and storage management. This method is particularly suitable for budget-constrained small clusters and small to medium-sized enterprises, providing a cost-effective and practical solution.

\section{GNN for Parallelism Optimization in classical topology}
\begin{figure*}
    \centering
    \includegraphics[width=0.75\textwidth]{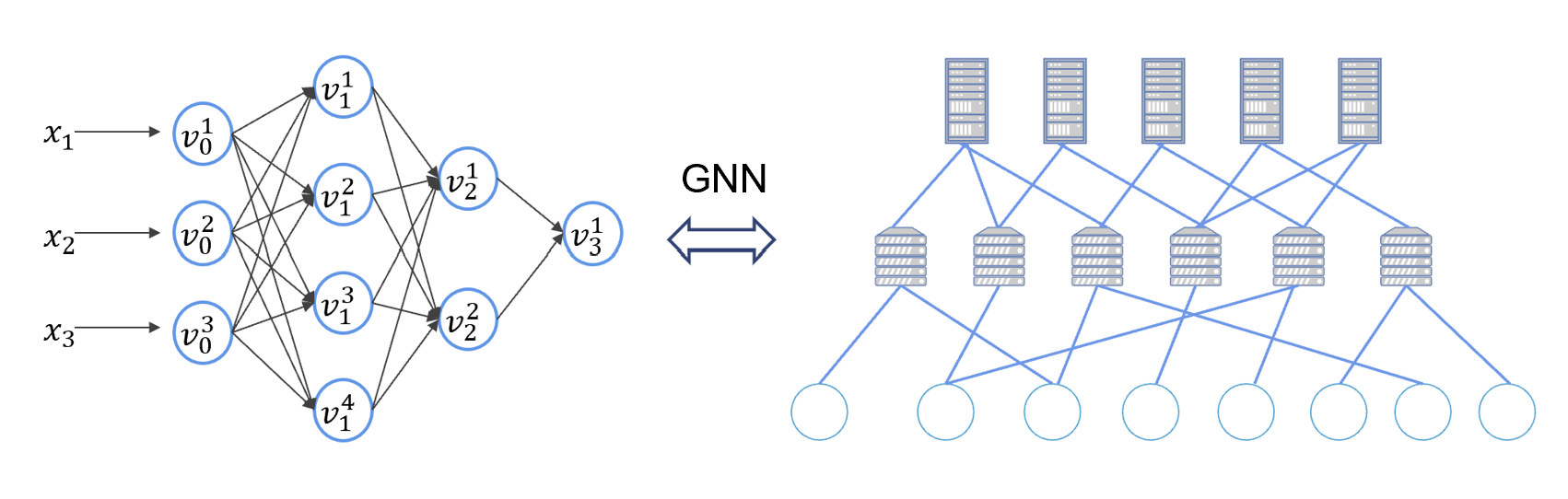}
    \caption{GNN for Network and Nerual Network}
    \label{fig:enter-label}
\end{figure*}
The previous sections detailed the physical and logical arrangements of CPUs, GPUs, and network interfaces, which fundamentally shape the performance characteristics of computing clusters. The use of GNNs in this section considers these physical connections and models them as graphs, allowing for more sophisticated analysis and optimization strategies based on network topologies.

Graph Neural Networks (GNNs) play a significant role in optimizing parallelism in neural networks and high-performance computing. We can unify the analysis of intra-node and inter-node topologies using the following steps and utilize minimum communication time to describe the effectiveness of different parallel levels.

\subsection*{Step 1: Establish the Graph Structure}
First, we consider the intra-node and inter-node topology as a graph \( G(V, E) \), where:
\begin{itemize}
    \item \( V \) is the set of nodes
    \item \( E \) is the set of edges representing connections between nodes
\end{itemize}

\subsection*{Step 2: Construct the Adjacency Matrix}
The adjacency matrix \( A \) of the graph \( G \) is defined as follows:
\[
A_{i,j} = 
\begin{cases} 
1 & \text{if there is an edge between node } i \text{ and node } j \\
0 & \text{otherwise}
\end{cases}
\]

\subsection*{Step 3: Define Node Neighborhood}
For each node \( v \), define its neighborhood \( N(v) \) as:
\[
N(v) = \{ u \in V \mid (v, u) \in E \}
\]

\subsection*{Step 4: Define Network Traffic under Parallel Levels}
For each node \( e \), define the network traffic under a certain parallel level as \( T(i) \):
\[
T(i) = \{ P_1, P_2, \ldots, P_n \}
\]
where \( P_n \) represents the network traffic of node \( e \) under parallel level \( n \).

\subsection*{Step 5: Describe the Effect of Parallel Levels}
Use the minimum communication time to describe the effectiveness of a certain parallel level. For each parallel level \( P_n \), we can compute its communication time \( T_{comm}(P_n) \).

\subsection*{Step 6: Construct the Time Dimension Matrix}
Embed the communication times of different parallel levels into the intra-node and inter-node topology graph to construct a matrix over the time dimension. Each element of this matrix represents the communication time at a specific parallel level and time point.

Assume we have two parallel levels \( P_1 \) and \( P_2 \), and we compute their communication times \( T_{comm}(P_1) \) and \( T_{comm}(P_2) \). The constructed time dimension matrix may look like this:
\[
T = \begin{bmatrix}
T_{comm}(P_1, t_1) & T_{comm}(P_1, t_2) & \ldots \\
T_{comm}(P_2, t_1) & T_{comm}(P_2, t_2) & \ldots \\
\end{bmatrix}
\]

This time dimension matrix can be used to analyze the efficiency of different parallel levels and optimize parallelism in neural networks.

\section{Future Work}
\subsection{Development of Task-Specific Blueprints}
In our future work, we will work on suitable blueprints for different distributed computing and machine learning training and inference tasks and develop reproducible measurement setups that will allow for the comparison of different architectures and configurations.
\subsection{Parallelism Optimization}
We will focus on optimizing parallelism to enhance computational efficiency. By analyzing and optimizing the parallel execution of different tasks in distributed systems, we can reduce communication overhead and improve the utilization of computational resources. Specifically, we will investigate how to maximize parallelism while maintaining task accuracy, thereby significantly improving overall system performance.
\subsection{Unified Analysis Framework}
To systematically analyze and optimize the performance of distributed computing systems, we plan to develop a unified analysis framework. This framework will integrate different types of data flows, communication patterns, and computational tasks, providing a comprehensive perspective for evaluating the system performance. Through this unified analysis, we can better identify bottlenecks and formulate targeted optimization strategies.

\section{Conclusion}
This study provides a robust guideline for academic researchers to navigate the complex landscape of computing resources. Integrating cost considerations with technical optimizations offers a strategic pathway to harness maximum computational power within constrained budgets, ultimately facilitating more advanced research and development activities. By considering the intra-node and inter-node topologies as a graph structure and using Graph Neural Networks to analyze the communication times of different parallel levels, we can effectively optimize parallelism in neural networks and high-performance computing. This method not only unifies the analysis of intra-node and inter-node topologies but also provides an effective framework for minimizing communication time.

In our future work, we will develop task-specific solutions for various distributed computing and machine learning tasks to enhance our framework for optimizing different architectures, thereby further strengthening the robustness of computational research. Additionally, we will focus on optimizing parallelism to improve computational efficiency, particularly by reducing communication overhead while maintaining task accuracy. We will also concentrate on creating a unified analysis framework that integrates different data flows, communication patterns, and computational tasks. This comprehensive framework will provide a systematic approach to identifying bottlenecks and formulating targeted optimization strategies, thereby further improving the performance of distributed computing systems.

\section{Acknowledgments}

This work was supported in part by the Guangzhou Municipal Key Laboratory on Future Networked Systems (024A03J0623), by the Guangdong Provincial Key Lab of Integrated Communication, Sensing and Computation for Ubiquitous Internet of Things (2023B1212010007), and by the Guangzhou Municipal Science and Technology Project 2023A03J0011. 

\dirk{some discussion about future trends?}
\dirk{discuss some ideas for use cases?}
\dirk{discuss some ideas for future work?}

\end{document}